\documentclass[5p,twocolumn]{elsarticle}

\usepackage{lineno,hyperref}
\modulolinenumbers[5]

\journal{Physica A}

\usepackage{inputenc}
\usepackage[english]{babel}
\usepackage{amsmath}
\usepackage{graphicx} 

\usepackage{natbib}

\begin{document}

\begin{frontmatter}

\title{Periodic spiking by a pair of ionic  channels
\tnoteref{t1}
}
\tnotetext[t1]{\textcopyright 2018. This manuscript version is made available under the CC-BY-NC-ND 4.0 license http://creativecommons.org/licenses/by-nc-nd/4.0/}

\author{L. Ram\'irez--Piscina\fnref{mycorrespondingauthor}}
\fntext[mycorrespondingauthor]{Corresponding author}
\ead{laure.piscina@upc.edu}
\address{Departament de F\'isica, Universitat Polit\`ecnica de Catalunya,\\
Avinguda Doctor Mara\~n\'on 44, 08028 Barcelona, Spain}
\author{J.M. Sancho}
\address{Departament de F\'isica de la Mat\`eria Condensada, Universitat de Barcelona,\\
Universitat de Barcelona Institute of Complex Systems (UBICS),\\
Mart\'i i Franqu\'es 1, 08028 Barcelona, Spain}
\date{\today}

\begin{abstract}
{Neuronal cells present periodic trains of localized voltage spikes involving a large amount of different ionic channels. A relevant question is whether this is a cooperative effect or it could also be an intrinsic property of individual channels. Here we use a Langevin formulation for the stochastic dynamics of a pair of Na and K ionic channels. These two channels are simple gated pore models where a minimum set of degrees of freedom follow standard statistical physics. The whole system is totally autonomous without any external energy input, except for the chemical energy of the different ionic concentrations across the membrane. As a result it is shown that  a unique pair of different ionic channels can sustain membrane potential periodic spikes. The spikes  are due to the interaction between the membrane potential, the ionic flows and the dynamics of the internal parts (gates) of each channel structures. The spike  involves a series of dynamical steps being the more relevant one the leak of Na ions. Missing spike events are caused by the altered functioning of specific model parts. The time dependent spike structure is comparable with experimental data.
}
\end{abstract}

\end{frontmatter}

\section{Introduction}
Neurons  exhibit a great variety of firing electrical patterns. It is recognized that the neuronal electrical activity depends, besides from the synaptic inputs, on its electrophysiological membrane properties of  the specific type of neuron \cite{Hille,Hammond}. 
These membrane properties ultimately depends on physical processes, such as the movement of ions through the molecular channels, the membrane potential dynamics, and the gating dynamics of the channels. 
These processes are indeed complex, involving a large hierarchy of biomolecular structures  from the atomic to the cellular and multicellular scales. 

However,
from the point of view of physical modeling, it is interesting to explain these processes by using a reduced formulation, with only a minimum of relevant physical mechanisms and variables.  One could then address for instance the question of whether the observed firing behavior could appear in a very simple device or it is instead necessary a large biological complexity or a whole collectivity of channels \cite{Hammond,fraser2007}.
Also, due the the large variety and complexity of real neural outputs, it is fundamental the recognition of the different fundamental firing patterns. Then one could in principle try to reproduce the most basic response patterns by using only the necessary ingredients in a unified framework, and to study the changes of these basic patterns when varying the physical parameters. Once this aim is achieved, it is open the possibility of introducing more elements in order to reproduce more complex patterns. This objective implies the choice of a simple theoretical scenario as starting approach.

Most theoretical approaches for action potential dynamics are based on the classic Hodking-Huxley \cite{H-H,Izhikevich} framework. It originally consists of deterministic equations for the dynamics of membrane permeability. 
At the level of individual channels other computational approaches consider microscopic details at the atomic scale by means of molecular dynamics simulations \cite{furini13}. Also mechanical models for the gating dynamics for a K channel have been proposed \cite{wawrzkiewicz2017impact}.
In the last years there has been an increasing interest on the role of channel noise in neural firing patterns \cite{tuckwell2005time,ozer2009controlling,ozer2009spike,sun2011effects,guo2016regulation,yu2017stochastic,maisel2017}.
Fluctuations have usually been modeled by using either master equations for the gate states 
\cite{goychuk2003,groff2009markov}, or by including stochastic terms into the membrane conductivities \cite{goldwyn2011}. Also the diffusion of ions inside the channel has been considered \cite{lee2002ion,pawelek2010asynchronous}.

Recently, a semi-microscopic approach for the stochastic dynamics of individual molecular channels was formulated \cite{ramirez1,ramirez2}. This approach uses only some relevant physical mechanisms acting on a minimum model. Variables representing the relevant degrees of freedom (ion positions and gate states) interact through a single energy functional, and a Langevin dynamics is then constructed by following standard rules from statistical physics. It is worth to remark the physical consistency of the resulting formulation. On the one hand the magnitude of the fluctuations verifies the fluctuation-dissipation theorem. On the other hand the working of the channel is autonomous, with the energy source being the chemical energy from the different ionic concentrations inside and outside the cell.
The resulting model was able to reproduce the basic properties of Na and K channels \cite{ramirez2} and the excitable properties of a single Na channel in the presence of K leak \cite{ramirez1}.

Here we will show how this approach, applied to a unique pair of channels following known physical mechanisms acting on the membrane, is able to generate a periodic firing pattern of electrical activity. We will find the effects of the external control parameters, and answer questions on how to control the stability of the periodicity or where are the sources of misfunctioning.

\begin{figure}[!ht]
\includegraphics[width=0.85\columnwidth,clip]{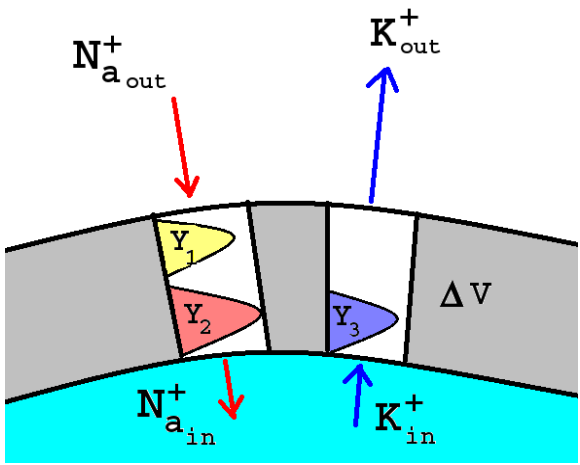}
\caption{Picture of the pair Na and K channel models with their respective gates $Y_1$, $Y_2$, $Y_3$, ionic concentrations $[Na^+]$, $[K^+]$, ionic fluxes (arrows), and membrane potential $\Delta V$.}
\label{Na-K-model}
\end{figure}

The structure of this paper is as follows. In the next section we will show how a couple of gating pores,  modeling Na and K channels, exhibit periodic firing with a well controlled period. Next we will study the changes in the response pattern by changes of different channel parameters and external conditions. We end with some conclusions. Some specific details of the approach are presented in the Appendix.

\section{The periodic spiking minimum device}

We have employed a minimum device consisting of a couple formed by two channel models,  as a rude simplification of Na and K voltage-gated ionic channels,  as pictured in Fig. \ref{Na-K-model}. Each model is a simple physical pore with active gates controlled by the membrane potential.
The Na-like  channel has two gates, 
whereas the K-like  channel has a single gate.
This physical structure has been designed \cite{ramirez1,ramirez2} to follow experimental observations on individual Na and K channels (see for instance Chapter 4 in Ref. \cite{Hammond}).

The physical variables of the model are the ion positions $x_i$,  the gate coordinates $Y_j$, and the membrane potential $\Delta V$. The $Y_i$ are bistable variables such that the value 0/1 corresponds to the corresponding gate being closed/open. Gates $Y_1$, $Y_2$ of the Na channel are activating and deactivating ({\it i.e.} open and close with membrane depolarization) respectively. According to experimental evidence, they are mediated by different parts of the channel structure \cite{Hammond,catterall2012voltage}. Gate $Y_3$ of the K gate is activating. 
These gates ($Y_1$, $Y_2$, $Y_3$) can be related to the activating ($m(\Delta V)$,  $n(\Delta V)$) and deactivating ($h(\Delta V)$) functions  appearing in the H--H formulation \cite{H-H,Izhikevich}.

We have considered a single degree of freedom ($Y$ variable) for each gate. Generalization to more $Y$ variables to account for the existence of several  voltage-sensing domains in real channels is straightforward, but it has not been considered here in order to keep the formulation of the model to a minimum.
The control parameters are the out cell ionic concentrations with fixed intra cell concentrations. Thus we implicitly neglect changes in bulk concentrations originated by the small local flow from the pair of channels.  

The model is defined by an energy functional $U$ describing the interaction between the different physical variables. The explicit construction of this functional is described in the Appendix. Following basic statistical physics the variables $x_i$, $Y_j$ follow a brownian stochastic dynamics \cite{ramirez1}:
\begin{eqnarray}
\gamma_x {\dot x_i} &=& - \partial_{x_i} U(x_i, Y_j, \Delta V) + \xi_i(t),
\label{eqx}\\
 \gamma_{Y_j}{\dot Y_j} &=&  - \partial_{Y_j} U(x_i, Y_j, \Delta V) +  \xi_{Y_j}(t), 
 \label{eqY}
\end{eqnarray}
where thermal noises fulfill
\begin{equation}
\langle \xi_a (t) \xi_b(t') \rangle = 2 \gamma_a \,k_B T\, \delta_{a,b}\, \delta (t- t'),
\end{equation}
and $\gamma_a$ are the corresponding frictions. Note that by construction the model verifies fluctuation-dissipation theorem. 
An analogous formulation for the coupling between ion and channel state, consistent with statistical physics, was already used in Ref. \cite{lee2002ion} to describe the stochastic behavior of singly occupied ion channels.

The numerical simulation of these equations, by a first order algorithm,  allows to record  the state of the gates and the position of the ions. The ionic concentrations in the bulk are implemented as boundary conditions at both ends of the channel for the Langevin dynamics of ions \cite{ramirezboundary}.
Finally, the dynamics of the membrane potential follows  the classical capacitor  equation
\begin{equation}
C_M \frac{d \Delta V}{d t} =  - \sum_i I_i,
\label{capacitorHH}
\end{equation}
where $C_M$ is the membrane capacity assumed to be constant and the r.h.s term includes all the ionic fluxes either across the channel or membrane leaks.
%
%

\begin{figure}
\includegraphics[width=0.85\columnwidth,clip]{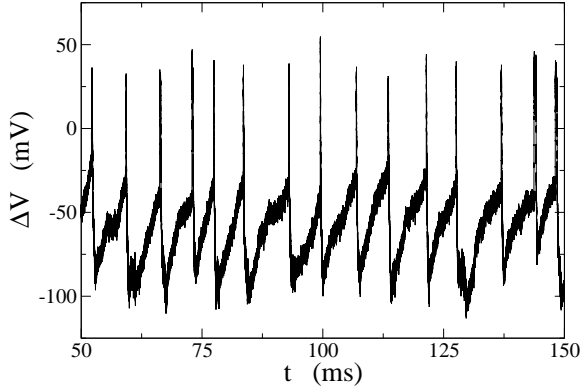}
\caption{Train of spikes with good periodicity and the same qualitative structure.  $[\text{K}_\text{out}]$ = 0.149 M.}
\label{fig.9kT-036}
\end{figure}

Letting the system evolve according the dynamical equations, without any external perturbation,   we observe that the membrane potential  present periodic pulses or spikes. We show a temporal evolution of this quantity in Fig.~\ref{fig.9kT-036}, which is very similar to those presented  in some experiments, as it is seen for instance in Chapter 17 in Ref. \cite{Hammond}.

The working of the system during a few spikes,  in this range of parameters, can be seen in Figs.~\ref{fig.puertas}, \ref{fig.puertas-zoom}, where the evolution of the three gate variables and the membrane potential are shown.  

\begin{table}[!htbp]
\begin{center}
\begin{tabular}[c]{|l|l|l|l|l|}
\hline
 Regime &   step   & $Y_1$ & $Y_2$   & $Y_3$  \\
\hline
Stand-by & \; a & $\;0$ & $\; 1$ & $\; 0$  \\
\hline
Pulse & \; b    &  $\; 1$  &  $\; 1$ &  $\; 0$  \\
\hline
Pulse & \; c  & $\; 1$  & $\; 1$ &  $\; 1$  \\
\hline
Refractory & \, $  d_1$ & $\; 1$  & $\; 0$ &$\; 1$ \\
\hline
Refractory & \, $ d_2$ & $\; 0$  & $\; 0$ &$\; 1$ \\
\hline
Refractory & \, $ d_3$ & $\; 0$  & $\; 1$ &$\; 1$ \\
\hline
\hline
Error  $Y_2$ & \, $b'$ & $\; 1$  & $\; {\bf 0}$ & $\; 0$ \\
\hline
Error  $Y_2$ & \, $c' = d_1$ & $\; 1$  & $\;{\bf 0} $ &$\; 1$ \\
\hline
Error  $Y_3$ & \, $ a' = d_3$ & $\; 0$  & $\; 1$ &$\;{\bf 1} $ \\
\hline
\end{tabular}
\end{center}
\caption{Table with the different regimes, steps and gate states, during the pulse. For the cases of missing spikes, the altered gate steps are also specified in boldface.}
\label{steps}
\end{table}

These figures reveal that any successful pulse involves three different regimes : stand-by, the spike event and the refractory state. These physical regimens are composed of steps (Table \ref{steps}):  stand-by from the membrane at the K-Nernst potential which corresponds to  the dynamical depolarizing (step a),  the spike (steps b and c) and  the restoring of the K-Nernst potential (step d). Being more explicit the description of each step is:

\begin{figure}[!htbp]
\includegraphics[width=0.85\columnwidth,clip]{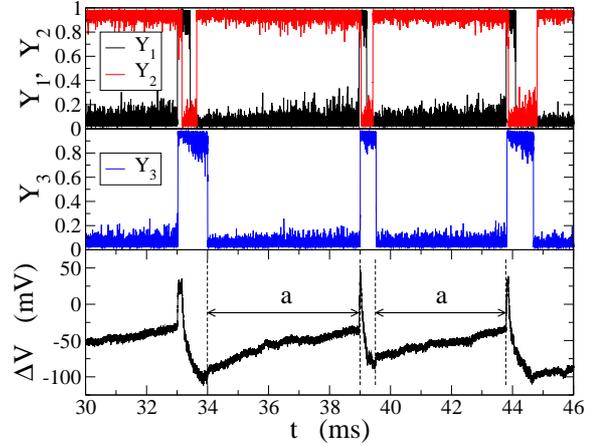}
\caption{Temporal evolution of Na and K channel variables  during three typical pulses. It can be seen that the dynamical step (a) is the main factor controlling the periodicity of the train of spikes. (Top) Na gate variables $Y_1$ (black line) and $Y_2$ (red line)  versus time; (Middle) gate variable $Y_3$ of the K channel; (bottom) membrane action potential.}
\label{fig.puertas}
\end{figure}

\begin{figure}[!htbp]
\includegraphics[width=0.85\columnwidth,clip]{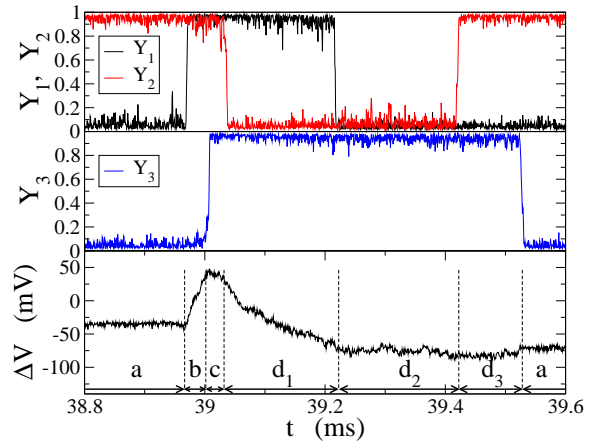}
\caption{Magnified time evolution during a single typical pulse (middle peak of Fig.~\ref{fig.puertas}). The shorter dynamical steps (b, c, d) specified in Table 1 are  shown in more detail. (Top) Gates variables $Y_1$ (black line) and $Y_2$ (red line) of the Na channel versus time; (Middle) gate variable $Y_3$ of the K channel; (bottom) membrane action potential.}
\label{fig.puertas-zoom}
\end{figure}

\begin{figure}[!htbp]
\includegraphics[width=0.85\columnwidth,clip]{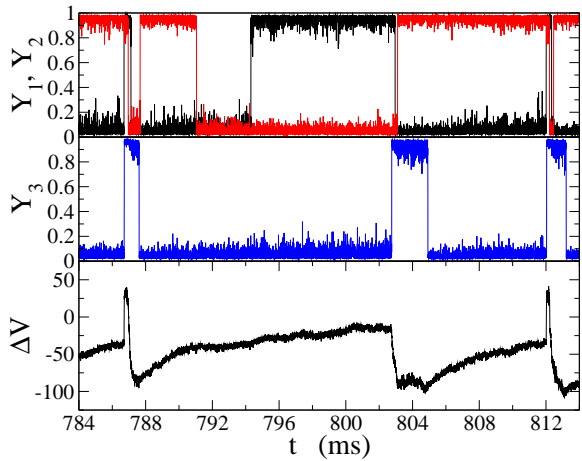}
\caption{Example of gate dynamics during a missing pulse close to $t=803$ ms due to an altered Na gates sequence ($Y_2$ closes before opening of $Y_1$). Color code as in previous figures.}
\label{fig.errorNa}
\end{figure}

\begin{figure}[!htbp]
\includegraphics[width=0.85\columnwidth,clip]{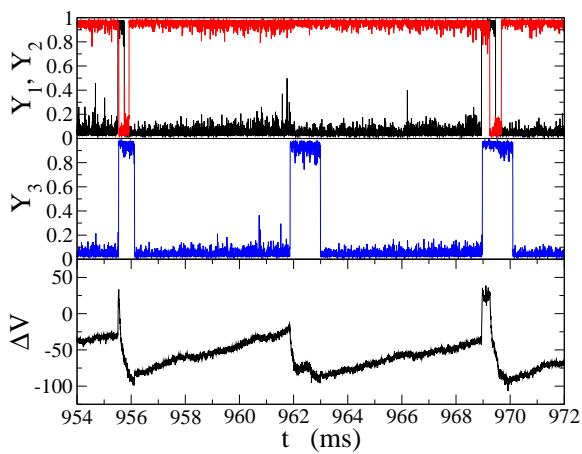}
\caption{Example of gate dynamics during a missing pulse close to $t=962$ ms due to altered K gate opening ($Y_3$ opens before $Y_1$). Color code as in previous figures.}
\label{fig.errorK}
\end{figure}

\begin{figure}[h]
\includegraphics[width=0.85\columnwidth,clip]{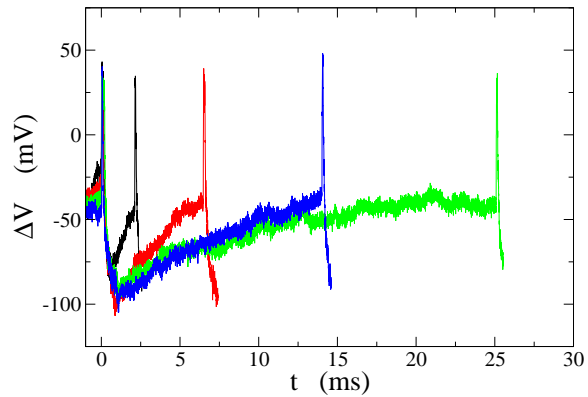}
\caption{Time evolution of the membrane potential between two consecutive pulses for four values  of the  $Y_1$ barrier height considered in Table~\ref{table-periods}. Times are shifted in order to place the first pulse of each case at $t=0$ and in this way to highlight the corresponding period. The results illustrate how the period increases with barrier height.  Black line: $V_d({Y1}) =$ 8 k$_B$T, red line: $V_d({Y1}) =$ 9 k$_B$T, blue line $V_d({Y1}) =$ 9.85 k$_B$T,  and green line: $V_d({Y1}) =$ 10 k$_B$T. }
\label{fig.comp-barrera}
\end{figure}

a).- This is the main dynamical process to trigger the pulse and to control the spike period. This step has  the largest time span where the Na gate  $Y_1$  is closed and  the $Y_2$ gate is open. The K gate $Y_3$  is just closed and the membrane potential is  the K Nernst value $\Delta V \sim  -92$ mV (See Fig. \ref{fig.puertas} at $t=34\,$ms). Now the $Y_1$ gate, although closed,  has nevertheless a small leak of Na ions. 
In Fig. \ref{fig.puertas} we see how this leak increases the action potential depolarizing the membrane. An analogous phase, the so called pacemaker depolarization, is found in Na channels of Purkinje cells \cite{Hammond}.

b) and c).- The pulse.  This regime is the shortest time interval, which is magnified in Fig. \ref{fig.puertas-zoom}. The Na leak  increasing  $\Delta V$ triggers the sequence of events in the Na channel that originates first the opening of gate $Y_1$, which permits a larger flux of Na ions, depolarizing the membrane by increasing  $\Delta V$ even further forming the pulse (b). Very quickly the gate $Y_3$ of the K channel also opens  and gate $Y_2$ of the Na channel closes. Then the flux of K ions towards the exterior of cell leads the membrane potential to reset its polarization, leading to more negative values of $\Delta V$ (c).


d).- Na gate $Y_2$  closes and  the membrane potential relaxes towards the K-Nernst potential   because during all this step the Na channel is closed and the K channel is open. In this regime the Na channel presents three different configurations of closed channel. d$_1$: $Y_1$ open and $Y_2$ closed, d$_2$: with  $Y_1$  and $Y_2$ both closed, and d$_3$ with $Y_1$ closed and $Y_2$ open. Then the Na channel rests in stand-by regime, until the K gate $Y_3$ closes and the depolarization starts a new cycle.

In this dynamics some cycles are observed to be missing because of the inherent stochasticity of the gate dynamics.
In the first example the Na door $Y_2$ closes before the completion of the  depolarization of the action potential. This misses a possible spike at $t=803$ ms (Fig. \ref{fig.errorNa}).  In the second example the K door  $Y_3$ opens reducing the membrane potential without a pulse at $t=962$ ms (Fig. \ref{fig.errorK}). The system goes back to the state $d_3$.
These two cases of missed spikes disturb the regularity of the train of spikes,  increasing the mean and the variance of the period. All these steps are summarized in Table \ref{steps}.

\section{Effect of the Na ions leak}

The pulse of the action potential is produced by the synchronized opening and closing of gates of the Na channel, and it is completely similar to the process described for the model of excitable membrane of Ref.~\cite{ramirez1}. The periodicity of the oscillatory behavior is controlled by the slow leak of the Na ions when both channels are closed, which introduces its time scale into the process. This slow leak could be produced by other processes or channels in more complex devices.
In our case the period of the oscillations should directly be related to the rate at which ions can cross the potential barrier of the closed gate $Y_1$. As it is well known the time scale for crossing a barrier is given by the Arrhenius (or Kramers) law, according to which $T\sim \exp V_d/k_BT$ for $V_d \gg k_BT$, where $V_d$ is the height of the barrier (in this case that of the $Y_1$ gate). To test this prediction we have performed simulations with different values of $V_d({Y_1})$. Results of mean periods, measured as the time intervals between membrane potential peaks, are presented in Table~\ref{table-periods}.
\begin{table}[!htbp]
\begin{center}
\begin{tabular}[c]{|l|l|l|l|l|}

\hline
   $V_d({Y_1}) $ (k$_BT$) &  \; 8   & \; 9 & \; 9.85   & \; 10  \\
\hline
Number of 
oscillations & \; 501 & \; 274 & \; 115 & \; 94 \\
\hline
Mean period $\langle T \rangle$  (ms) & $\; 4.03 \; $  &  $\; 7.33 \; $  &  $\; 17.3 \; $  &  $\; 20.9$  \\
\hline
Std dev. $\sigma_T$ (ms) & $\; 2.77$  & $\; 3.33$  & $\;\; 7.2$ &  $\; 10.2$  \\
\hline
\end{tabular}
\end{center}
\caption{Oscillation periods for different $Y_1$ barrier heights, calculated for a simulation time span of 2025 ms.}
\label{table-periods}
\end{table}

Note that according to the Kramers' law, by increasing the barrier height in 1$\,k_B$T the crossing time should increase in a factor $e=2.718...\,$ for $V_d \gg k_BT$. According the results in Table~\ref{table-periods}: $T_{9\text{$k_B$T}}/T_{8\text{$k_B$T}}=1.8$, 
$T_{10\text{$k_B$T}}/T_{9\text{$k_B$T}}=2.8$.
While the order of magnitude is correct, the result is better for the larger barrier case as expected. In Fig. \ref{fig.comp-barrera} four examples with different periods are plotted.

\section{Effect of external cell  concentrations}

For a specific Na-K pair of channels, their internal parameter values, including concentrations,  are fixed, so the most viable possibility of some external control is to change the external (out cell) Na and K concentrations.

\begin{figure}[ht]
\includegraphics[width=0.85\columnwidth,clip]{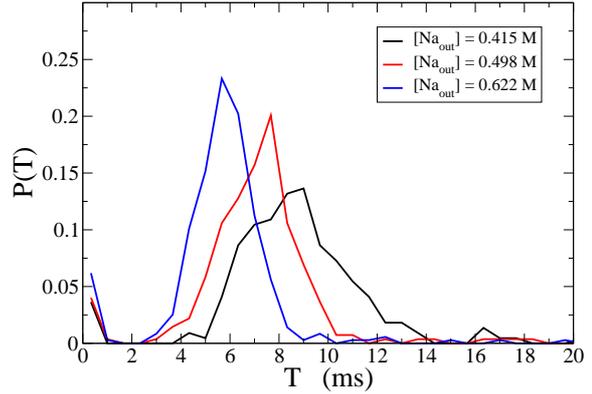}
\caption{Distribution $P(T)$ of periods for Na$_\text{out}$  concentrations  of Table \ref{table-periods-Na}, which are indicated in the inset. Periods are shorter for larger Na$_\text{out}$ concentration, due to an increase of Na leak.}
\label{histograma-Na}
\end{figure}

{\bf Na$_\text{out}^+$ concentration:}
We expect that the height of the spike follows  the Na concentration because the Na Nernst potential also increases as $\ln[\rho_{Na}(out)/\rho_{Na}(in)]$. Moreover the period of the oscillations is decreased because of the leak increase. Numerical simulation for different Na external concentrations were conducted to confirm this prediction. The distribution of periods can be seen in Fig. \ref{histograma-Na} and mean and variance values in  Table \ref{table-periods-Na}.

\begin{table}[!htbp]
\begin{center}
\begin{tabular}[c]{|l|l|l|l|l|}
\hline
 $\rho_{Na}$(out) (M)  &  \;\; 0.415  & \;\; 0.498  & \;\; 0.622  \\
\hline
Number of 
oscillations & \;\; 220 & \;\; 274 & \;\; 356  \\
\hline
Mean period $\langle T \rangle$ (ms)  & $\;\; 9.10$  &  $\;\; 7.33$  &  $\;\; 5.65$   \\
\hline
Std dev. $\sigma_T$ (ms) & $\;\; 5.38\;\;$  & $\;\; 3.33\;\;$  & $\;\; 2.14\;\;$   \\
\hline
\end{tabular}
\end{center}
\caption{Mean and variance of spike periods  for different external Na concentrations, calculated for a simulation time span of 2025 ms.}
\label{table-periods-Na}
\end{table}

\begin{figure}[ht]
\includegraphics[width=0.85\columnwidth,clip]{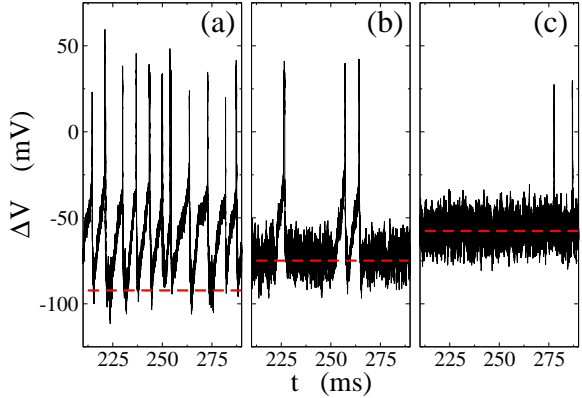}
\caption{Three typical time evolutions of membrane potential for different K$_\text{out}$ concentrations.
Periodicity is lost for higher K$_\text{out}$ concentration.
$[\text{K}_\text{out}]=$ (a)  0.208 M, (b) 0.415 M, (c) 0.830 M. The red dashed line indicates the  Nernst potential corresponding to each concentration.}
\label{fig.9kT-varios-k}
\end{figure}

\begin{figure}[!htbp]
\includegraphics[width=0.85\columnwidth,clip]{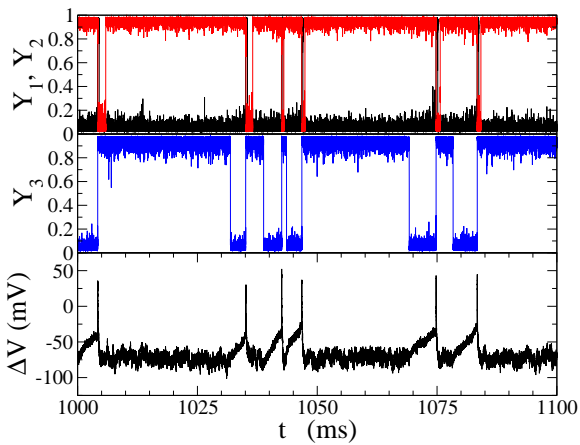}
\caption{Gate dynamics for the case (b) of Fig.~\ref{fig.9kT-varios-k} ($[\text{K}_\text{out}] = $ 0.415 M. Periodicity is lost and spikes appear at random times as it corresponds to an excitable regime. Color code as in Fig.~\ref{fig.puertas}}
\label{fig.gates-rhoK-1}
\end{figure}

{\bf K$_\text{out}^+$  concentration:}
We have seen in simulations that the period is rather insensitive to moderate variations of the external values  of  K concentration. When increasing it further a new effect appears. For higher values of K$_\text{out}$ concentration the difference with the internal concentration is lower, and then with the channel K open the membrane polarization is lower (i.e. the potential reach smaller negative values), as it can be seen in Fig.~\ref{fig.9kT-varios-k}. As a result the K channel has less tendency to close. Then, for larger time spans,  the system remains in a state with the Na channel closed and the K channel open. 
In such state it is necessary a larger fluctuation to induce the closing of the K channel and initiate the depolarization of the membrane that will produce the spike (Fig.\ref{fig.9kT-varios-k}b).

\begin{figure}[!htbp]
\includegraphics[width=0.85\columnwidth,clip]{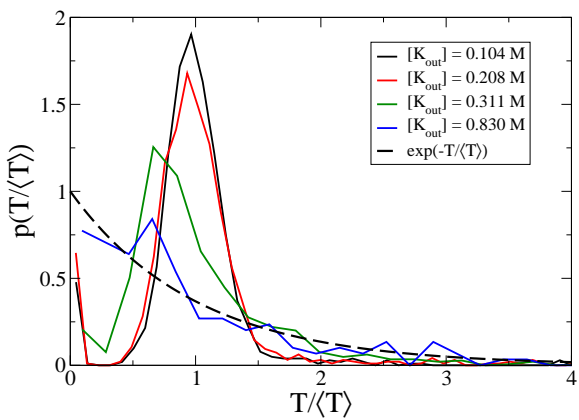}
\caption{Scaled distribution of  periods, for different values of the external K$_\text{out}$ concentration, indicated in the inset. Continuous lines: simulations of cases of Table~\ref{table-periods-rho}. The dashed line is the exponential distribution, which is characteristic of an excitable regime dominated by fluctuations.}
\label{histograma-k}
\end{figure}

This behavior can be seen in more detail in Fig.~\ref{fig.gates-rhoK-1} for a  $[\text{K}_\text{out}]= 0.415\,$M. Note that the periodicity of the spikes, given until now by the time scale of the Na leak, 
is now broken. Instead, to this time it is now added the waiting time for the stochastic closing of the K channel. This introduces a new source of stochasticity. 

For even larger  K$_\text{out}$ concentrations the effect  is stronger and the polarization of the membrane is weaker. As a consequence there are very few events of closing of the K gate. In this case some peaks of the membrane potential can be observed due to strong fluctuations, but now not necessarily associated with closings of the K channel (Fig.~\ref{fig.9kT-varios-k}c). 
%
%
This is then an excitable regime controlled by  fluctuations.
In this regime the apparition of peaks is expected to be a stochastic poisson process,
with an  exponential distribution of waiting times,  and then
a standard deviation equaling the mean value. In our case, for the lower  K$_\text{out}$ concentrations the standard deviations are consistently lower than the mean values, indicating the presence of a well defined period. However for the larger concentrations both values are very similar, as seen in Table~\ref{table-periods-rho}, indicating a poisson process. The period distributions, plotted in  
Fig. \ref{histograma-k}, clearly show the two limiting behaviors from an oscillatory regimen to an excitable one.

\begin{table}[!htbp]
\begin{center}
\begin{tabular}[c]{|l|l|l|l|l|l|l|}
\hline
 $\rho_K^\text{out}$(M) &  0.0415 & 0.104 & 0.311  & 0.21  & 0.415 & 0.83 \\
\hline
$N_\text{osc}$ & 1086 & 1115 & 1079 & 766 & 324 & 159 \\
\hline
$\langle T \rangle$ (ms) & 7.45 &  7.25 &  7.50 & 10.6 & 24.8 & 49.9 \\
\hline
$\sigma_T$ (ms) & 3.08 & 3.35 & 3.68 & 6.30 & 19.3 & 52.5 \\
\hline
\end{tabular}
\end{center}
\caption{Spike periods for different external K concentrations, calculated for a simulation time span of 8100 ms. Other data from Tables~\ref{table-param1} and  
\ref{table-param2}.}
\label{table-periods-rho}
\end{table}

\section{Conclusions}

We have shown that the periodic spiking activity of the action potential can be modeled by an unique pair of Na and K ionic channels, by using a minimal semi-microscopic approach \cite{ramirez1}.
The emergence of regular spikes is caused by the interaction between the channel gates, ionic flows and the membrane potential, within a unified physical framework.
The system is autonomous and it enters in the oscillatory regime without any kind  of perturbation or energy supply.  The only source of energy is the chemical energy of the ionic concentration differences between both sides of the membrane. This is assumed to be controlled by other cell mechanisms, such as molecular ionic pumps, not considered here.
The dynamical patterns of the spike events are very similar to those observed in experiments, with a clear stand-by regimen of membrane depolarization and a very narrow pulse.

Given the simplicity of the approach it has been possible to establish the relevant steps and involved variables in the periodic firing. Special attention has been paid to the external Na and K concentrations. In particular the K concentration has shown to be responsible for the transition from an oscillatory regime, with a well defined period, to an excitable regime, with stochastic waiting times. The period is controlled by the Na leak, which could be changed to model different neurons by using an internal parameter such as $V_d(Y_1)$. In our simulations we have obtained examples of spike intervals from $2$ ms to $25$ ms, but results are in principle not limited by these values. 

The approach uses basic physical processes, described with a minimum of degrees of freedom, implemented by using standard formalisms of statistical physics. This allows for further possibilities of including other physical elements and mechanisms, which should permit to cover additional specific firing patterns. They could further be combined to deal with more complex neural activity.

\section*{Acknowledgments}

This work was supported by the Ministerio de Economia y Competividad (Spain) and FEDER (European Union), under
projects FIS2015-66503-C01-P2/P3
and by the Generalitat de Catalunya Projects 2009SGR14 and 2014SGR878. 

\section*{Appendix: Model details}

\setcounter{equation}{0} 
\setcounter{table}{0} 
\renewcommand{\theequation}{A-\arabic{equation}}
\renewcommand{\thetable}{A\arabic{table}}

The total potential energy of the system is \cite{ramirez1}
\begin{eqnarray}
&&U(x_i, Y_j, \Delta V) = \\
&&  \sum_i V_i(x_i, \Delta V) + \sum_j V(Y_j, \Delta V) + 
 \sum_{i,j} V_I(Y_j,x_i).  \nonumber
 \label{pot}
\end{eqnarray}
This potential has 
three different terms corresponding to the interactions between elements.
The first one 
is the potential seen by the ions  inside the channel
\begin{equation}
V_i(x_i, \Delta V)= \frac{q\Delta V}{L} ( x_i- L), \qquad 0 < x_i < L,
\label{potmem}
\end{equation}
where $\Delta V$ is the potential difference between both sides of the membrane, and $L$ is the length of the channel.
The gate variables $Y_j$ evolve with the potential
\begin{eqnarray}
V(Y,\Delta V)&=& V_0 \left[ -a \ln (Y(1-Y)) - b (Y-0.5)^2 \right] 
\nonumber\\
 &+& Q(\Delta V - \phi_{\text{ref}}) Y,
 \label{potgate}
\end{eqnarray}
where the  first part represents the bistable internal structure. With $a\ll b$ it is a simple potential with two well minima  at the closed ($Y_j \sim 0$) and  open ($Y_j \sim 1$) states.
In the last term the parameter $Q$ is the  effective charge of the gate sensor and  $\phi_{\text{ref}}$ is the reference potential that determines the $\Delta V$ value at which both states are equally probable. The values for the different parameters 
(Table \ref{table-param2}) are characteristic of each specific channel and are chosen to enter into the experimental scales.

\begin{table}[h]
\begin{center}
\begin{tabular}[c]{|l|l|}
\hline
 $\gamma_{A}$ particle friction & $2 $ $\mu$s meV/nm$^2$ \\
 $\gamma_{B}$ particle friction & $200 $ $\mu$s meV/nm$^2$ \\
K$_B T$ &  $25$ meV  \\
$L$ channel length & $4\, $nm\\
$A$ channel section & $\, $4 nm$^2$\\
$c_0^{\text{A}}(in) $& $4.15$ mM\\
$c_0^{\text{B}}(in)$
& $8.30$ M\\
$C_M$  effective capacity & $1.25$ charges/mV \\
\hline
\end{tabular}
\end{center}
\caption{Physical fixed control parameter values used  in the simulations. Both ions have a positive charge $q=+1$ e.}
\label{table-param1}
\end{table}

\begin{table}[!htbp]
\begin{center}
\begin{tabular}[c]{|l|l|l|l|l|l|l|l|l|}
\hline
  & $\gamma$ &  $V_0$& $V_d$& $Q$& $\phi_{ref}$ & a & b &  $x_c$    \\
  & \; & \small{k$_B$T} &  \small{k$_B$T}&\small{e} & \small{mV} &    & & \small{nm}  \\ 
\hline
$Y_1$ &  1000 &   7 &   9  & +12  & -35 & 0.2 &  7 & 1.0  \\
$Y_2$ &  4000 &   7 &  10  & \; -8  & -35 & 0.2 &  9 & 3.0    \\
$Y_3$ &  4000 &   7 &   8  & +10  & -15 & 0.2 &  7 & 3.0    \\

\hline
\end{tabular}
\end{center}
\caption{Gates parameters of the Na and K channels. Units for $\gamma$ are $\mu$s\,meV/nm$^2$ and $\sigma= 0.283\,$nm.}
\label{table-param2}
\end{table}

Finally, the effect of the collisions of ions with the  gate  is modeled by the potential energy 
\begin{equation}
V_I(Y,x_i) = V_d f(Y)  \exp \left(-\frac{(x_i - x_c)^2}{2 \sigma^2} \right),
 \label{potint}
\end{equation}
where  $x_c$ is the  position of the center of the gate  inside the channel,  and $\sigma$ is its width. 
The height of the barrier is modulated by the function $f(Y)$, depending on the state (open or close) of the gate, with a maximum value $V_d$. For  the  envelope modulating  function the expression $f(Y) = (1 + \cos \pi Y)/2$ is used which 
 has the values $f(0) = 1$ for the close state, and $f(1)= 0$ for the open state. 
 The parameters of the  model are indicated in Tables ~\ref{table-param1}-\ref{table-param2}.

\bibliography{NaK.bib}


\end{document}